\documentclass[aps,twocolumn,amssymb,fleqn,showpacs]{revtex4} %fleqn: make all eqns left alighed
\usepackage{epsfig,amssymb,amsmath,graphicx,subfigure,hyperref }  % pdfpages and graphicx are added   hyperref(set hyperlink for content)

\newcommand{\be}{\begin{eqnarray}}   %%\nonumber\\
\newcommand{\ee}{\end{eqnarray}}

\begin{document}

\title{Crystalline Order on Catenoidal Capillary Bridges}
\author{Mark J. Bowick and Zhenwei Yao}
\affiliation{Department of Physics, Syracuse University, Syracuse,
New York 13244-1130, USA}

\begin{abstract}
We study the defect structure of crystalline  particle arrays on
negative Gaussian curvature capillary bridges with vanishing mean
curvature (catenoids). The threshold aspect ratio for the
appearance of isolated disclinations is found and the optimal
positions for dislocations determined. We also discuss the
transition from isolated disclinations to scars as particle number
and aspect ratio are varied.

\end{abstract}
\maketitle

Two-dimensional ordered phases of matter on  spatially curved
surfaces have several features not found in the corresponding
phase for planar or flat space systems~\cite{advphysreview}. For
crystalline order on surfaces of spherical topology where
disclination defects are required by the topology itself, Gaussian
curvature can drive the sprouting of disclination defects from
point-like structures to linear grain boundary scars which freely
terminate in the
crystal~\cite{interactingdefectsfrozen,science_sphere,BCNT_2002,BCNT_2006}.
Even for surfaces such as the torus which admit completely
defect-free crystalline lattices, the energetics in the presence
of Gaussian curvature can favor the appearance of isolated
disclination defects in the ground
state~\cite{torus1,torus_order}. For the axisymmetric torus with
aspect ratio between 4 and 10, isolated 5-fold disclinations
appear near the line of maximal positive Gaussian curvature on the
outside and isolated 7-fold disclinations appear near the line of
maximal negative Gaussian curvature on the
inside~\cite{torus-database}.  The ground states in these systems
are thus distinguished by a defect structure that would be
energetically prohibitive in flat space. It is certainly
worthwhile to explore as many settings as possible in which there
are qualitative changes in the fundamental structure of the ground
state, within a given class of order, purely as a consequence of
spatial curvature.

The richest  confluence of theoretical and experimental ideas in
the area of curved  two-dimensional phases of matter has been in
colloidal emulsion physics in which colloidal particles
self-organize at the interface of two distinct liquids, either in
particle-stabilized (Pickering)
emulsions~\cite{particlestabilizedemulsions,Pickering_1907}  or
charge-stablized emulsions~\cite{LBHSC:2007,LZRCB:2007}.
Two-dimensional (thin-shell) spherical crystals form at the
surface of droplets held almost perfectly round by surface
tension. The ordered configurations of particles may be imaged
with confocal microscopy and the particles manipulated with
optical tweezers~\cite{colloidosomes,science_sphere,toptweezers}.
Macroscopic examples of crystalline order on variable positive
Gaussian curvature surfaces have been constructed by forming a
soap bubble raft on a spinning liquid~\cite{BGST:2008} and the
nature of the order has been analyzed
theoretically~\cite{paraboloid_order}.

Glassy liquids on negative Gaussian curvature manifolds have also received considerable attention~\cite{RN:1983,TSV:2010,STN:2009,MK:2007,MK:2008}. The simplest such manifold conceptually is the constant (negative) curvature hyperbolic plane $H^2$ and it even appears that the hyperbolic plane can be isometrically embedded as a complete subset of Euclidean 3-space, although not differentiably~\cite{HT:2001}. Physical realizations of negative Gaussian curvature manifolds in condensed matter physics will almost always have variable Gaussian curvature. The inner wall of the axisymmetric torus ($S^1\times S^1$) has integrated Gaussian curvature equal to $-4\pi$, balancing an equal and opposite integrated Guassian curvature on the outer wall. This is responsible for the novel ground states noted above. Gaussian bumps have regions of both positive and negative Gaussian curvature and the minimal-type surfaces found in bicontinuous phases of amphiphilic bilayers have spatially extended variable Gaussian curvature that is negative on average~\cite{SC:1989}.

Recently crystalline particle arrays on variable Gaussian curvature surfaces has been studied by assembling particles on capillary bridges formed by glycerol in bulk oil spanning two flat parallel plates\cite{pleats}. Configurations may be imaged by confocal microscopy and even manipulated with laser tweezers. The interface between the inner fluid of the capillary bridge and the outer bulk fluid is a surface of revolution with a constant mean curvature (CMC) determined by the pressure difference between the two fluids~\cite{deGennes,gangwebsite}. Capillary bridges minimize the surface area at fixed volume and perimeter and appear in the classical work of Delaunay~\cite{eells,Delaunay}. The value of the mean curvature and hence the underlying surface may be changed by varying the spacing between the plates.

Three classes of the Delaunay surfaces have negative Gaussian curvature - the nodoids, unduloids and catenoids. We will focus on the most analytically tractable case of a catenoidal capillary bridge in which the mean curvature is everywhere zero. Capillary bridges themselves have wide-ranging application. They play an essential role in adhesion, antifoaming, the repelling coffee-ring effect and in the origin of attractive hydrophobic forces~\cite{anti_coffee_ring,capillary_bridge_review,capillary_bridge_review2,addcoffee}.

\begin{figure}[tbp]
\centering \subfigure[]{
\includegraphics[width=2.5in]{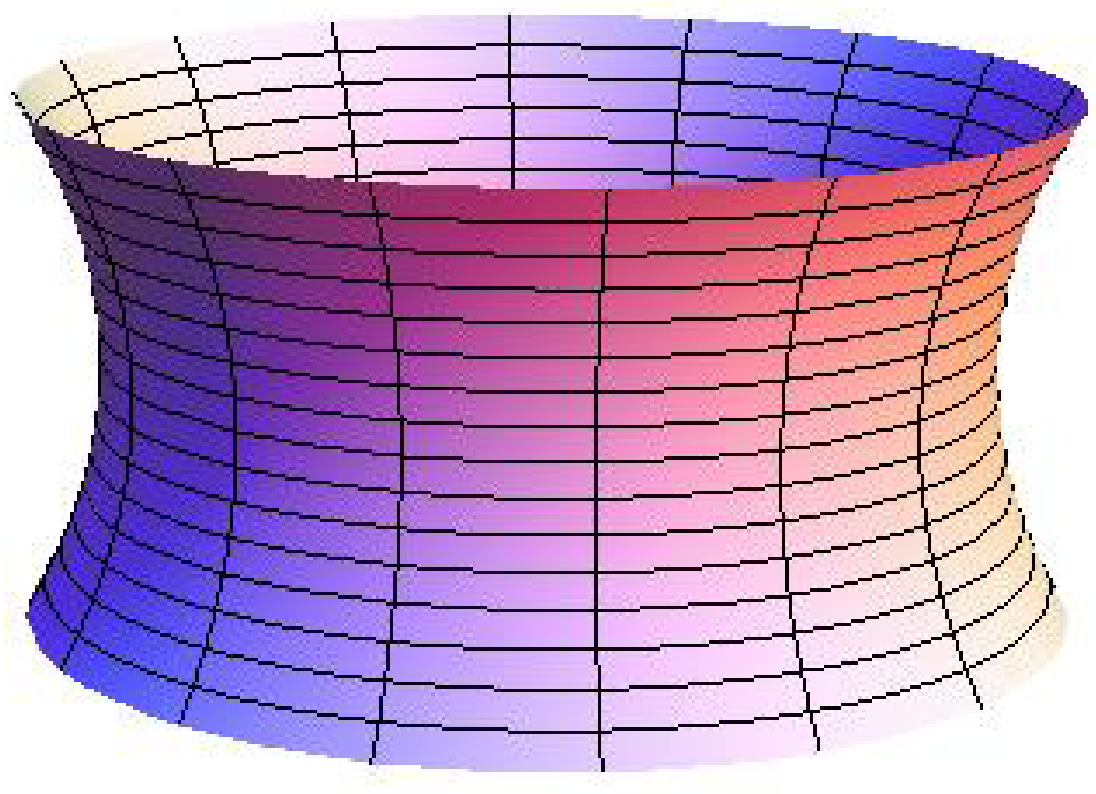}} %%,bb=37 16 400 249
\subfigure[]{
\includegraphics[width=2.2in,angle=-90]{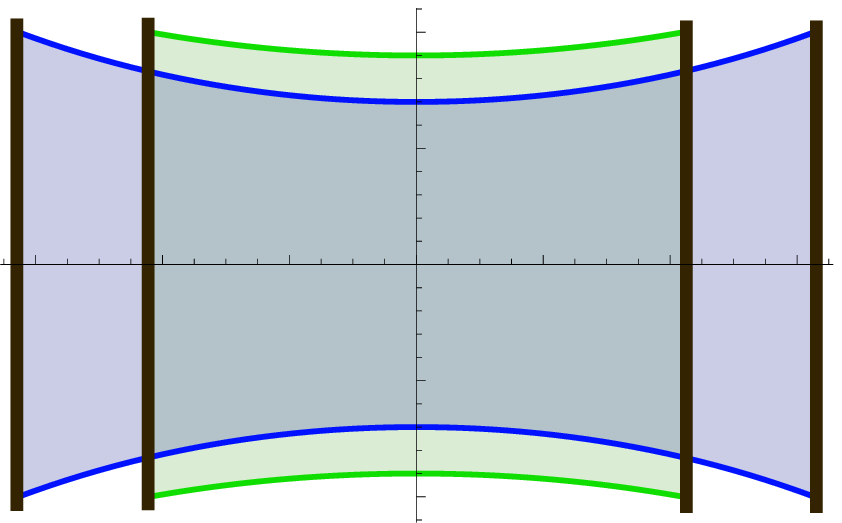}}
\caption{(a) The three dimensional shape of a catenoid with aspect
ratio $c=0.85$. (b) A catenoid with aspect ratio $c=0.9$ (green)
deforms to $c=0.7$ (blue). } \label{catenoid}
\end{figure}

The shape of a capillary bridge with mean curvature $H$ follows by solving \be
2H=\frac{-r''}{(1+r'^2)^{3/2}}+\frac{1}{r \sqrt{1+r'^2}} ={\rm const.}  \
, \label{Hrz}\ee where $ 2H\equiv \frac{1}{R_1}+\frac{1}{R_2}$,
$r=r(z)$ is the representation of a surface of revolution with symmetry axis z and $R_1$ and $R_2$ are the two principal radii of curvature at any point. The solution corresponding to the special case $H=0$ (a minimal surface) is
\be
\vec{x}(u,v) = \left(
\begin{array}{c}
c\ \cosh(\frac{v}{c})\cos u \\
c\ \cosh(\frac{v}{c})\sin u  \\
v \\
\end{array} \right),
\label{xcatenoid}\ee where $u\in [0,2\pi)$ and $v \in (-z_m,z_m)$.
This surface is the well-known {\em catenoid} parameterized by the
radius of the waist $c$ located in the  $z=0$ plane (see
Fig.\ref{catenoid}). From the non-zero metric components $ g_{uu}=
c^2 \cosh^2(\frac{v}{c})$ and $g_{vv}= \cosh^2(\frac{v}{c}) $,
one can obtain the Gaussian curvature  $ K\equiv \frac{1}{R_1}
\frac{1}{R_2}= -\frac{1}{c^2} {\rm sech}^4(\frac{v}{c})=-\frac{1}{g}$,
where g is the determinant of the metric tensor. We see explicitly
that the metric completely determines the Gaussian curvature. The
solution to Eq.($\ref{Hrz}$) for $H\neq 0$
is~\cite{capillarybridge1} \be \vec{x}(u,t) = \left(
\begin{array}{c}
\gamma \triangle(\theta,t) \cos u \\
\gamma \triangle(\theta,t) \sin u \\
\alpha+\gamma (E(\theta,t)+F(\theta,t) cos\theta) \\
\end{array} \right),
\label{unno}\ee where $t \in[t_0,t_1]$, $u\in [0,2\pi)$, $\
E(\theta,t)=\int_0^t \Delta(\theta, \tilde{t}) d \tilde{t} ,\
F(\theta,t)=\int_0^t 1/\Delta(\theta,\tilde{t}) d\tilde{t}$ and
$\Delta(\theta,t)=\sqrt{1-\sin^2\theta\ \sin^2 t} $. $\gamma$
plays the role of a scale factor. The curves generated are
periodic in $t$ with period $\pi$ and have maxima at $t=k \pi$ and
minima at $t=(k+1/2) \pi$ for integer $k$. The value of $\theta$
controls the shape of the profile: it generates an unduloid for
$\theta \in [0,\pi/2)$, a nodoid for $\theta \in (\pi/2,\pi)$; and
a semi-circle for $\theta=\pi$. The shapes of the capillary
bridges in the work of ~\cite{pleats} can be fit by
Eq.($\ref{unno}$), the general CMC expression with $H\neq 0$.

Here we study crystalline order on the simplest case of a
catenoidal capillary bridge ($H=0$) in the framework of continuum
elasticity theory~\cite{Nelson_France,interactingdefectsfrozen,advphysreview}.
For simplicity, we measure all lengths in units of the radius of
the contact disk.

The topology of the capillary surfaces we study is that of the annulus, with Euler characteristic zero, since the liquid bridge makes contact with the plates at the top and bottom.
Such a surface admits regular triangulations with all particles having coordination number 6. Although defects (non 6-fold coordinated particles) are not topologically required
they may be preferred in the crystalline ground state for purely energetic reasons since negative Gaussian curvature will favor the appearance of 7-coordinated particles (-1 disclinations).  To determine the preferred defect configuration we map the microscopic interacting particle problem to the problem of discrete interacting defects in a continuum elastic background. The defect free energy $F_{el}$, in the limit of vanishing core energies, may be expressed in the
form~\cite{paraboloid_order,advphysreview} \be
F_{el}=\frac{1}{2}Y\int_M
G_{2L}(\vec{x},\vec{y})\rho(\vec{x})\rho(\vec{y}) \
d^2\vec{x}d^2\vec{y} .\label{Eel}\ee Here
$G_{2L}(\vec{x},\vec{y})$ is the Green's function for the
covariant biharmonic operator on the surface $M$, $Y$ is the Young's
modulus for the crystalline packing, and $\vec{x}$ and $\vec{y}$
are position vectors on the surface. The effective topological
charge density is $\rho(\vec{x})= \frac{\pi}{3}q(\vec{x})-
\gamma^{ij} \nabla_j b_i(\vec{x})-K(\vec{x}) \label{rho3terms} $,
in which the first term is the disclination charge density
$q(\vec{x})=\sum_{\alpha} q_{\alpha}
\delta(\vec{x}-\vec{x}_{\alpha})$, the second term is the
dislocation density
$b_i(\vec{x})=\sum_{\beta}b^{\beta}_i\delta(\vec{x}-\vec{x}_{\beta})
$ and $K(\vec{x})$ is the Gaussian curvature. In this expression
$\gamma^{ij}=\epsilon^{ij}/\sqrt{g}$. By introducing $\chi$ and
$\Gamma$ such that $ \triangle^2 \chi(\vec{x})=Y \rho(\vec{x}) $
and $ \Gamma(\vec{x})=\triangle \chi(\vec{x}) $, Eq.($\ref{Eel}$)
can be written in a more compact form  \be F_{el}=\frac{1}{2Y}\int_M
\Gamma^2(\vec{x})\ d\vec{x},\label{Felchi} \ee in which
$\Gamma(\vec{x})/Y=\int G_L(\vec{x},\vec{y})\rho(\vec{y})\
d\vec{y}+U(\vec{x})  $ with $\triangle U(\vec{x})=0$ and
$G_{L}(\vec{x},\vec{y})$ is the Green's function for the covariant
Laplacian on $M$ which satisfies $ \triangle
G_L(\vec{x},\vec{y})=\delta(\vec{x},\vec{y}); \vec{x},\vec{y}\in M $ with
the boundary condition $ G_L(\vec{x},\vec{y})=0; \vec{x},\vec{y}\in
\partial_M $. By conformally mapping the surface parameterized by $\{ u,v \}$
onto an annulus in the complex plane via $z=\rho(u,v)e^{iu}$, the
Green's function $G_L$ is found to be \cite{paraboloid_order} $
G_L(\vec{x},\vec{y})=\frac{1}{2\pi}\ln|(\rho_0^{-1}
z(\vec{x})-z(\vec{y}))/(1-\rho_0^{-1} z(\vec{x})\bar{z}(\vec{y}))|
$, in which $\rho_0$ is the radius of the outer circle of the annulus
in the complex plane. For a catenoid, the conformal mapping is given
by $\rho(u,v)=c\ e^{|v|/c}$ and $\rho_0=c\ \exp(\textrm{arcsech}
(c))$.

\begin{figure}[htbp]
\centering\subfigure[]{
\includegraphics[width=2.3in]{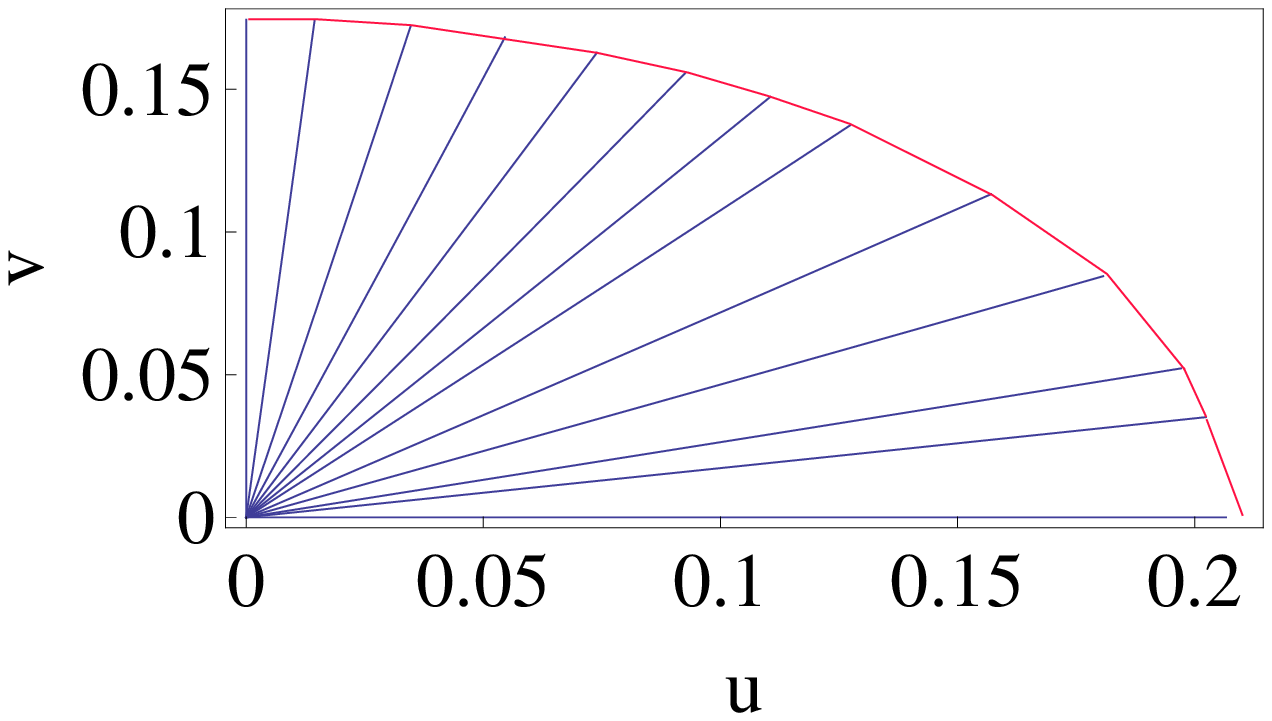}}
\subfigure[]{
\includegraphics[width=1.8 in]{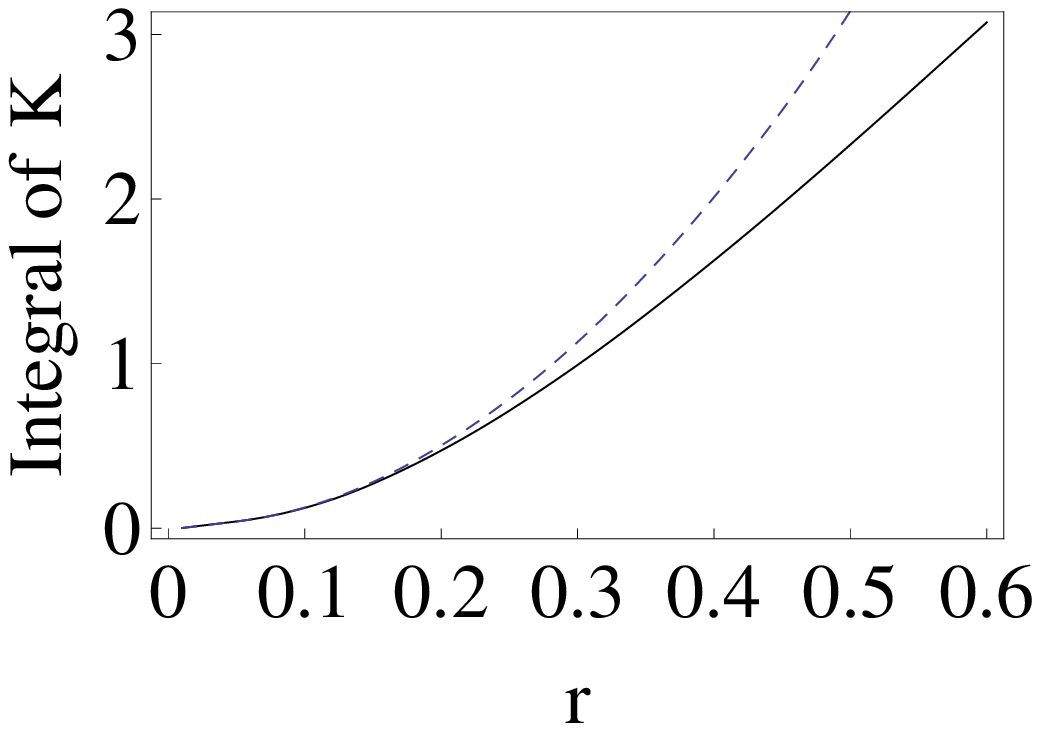}}
\caption{(a) A family of geodesics in $\{ u,v\}$ coordinates centered at a
point on the waist of a catenoid with $c=0.85$. (b) The $|K(0)| \pi r^2$
(dashed curve) and numerical result (solid curve) of the integrated
Gaussian curvature over a geodesic disk of radius $r$ versus $r$. $c=1/2$. }
\label{geodesics}
\end{figure}

Disclinations are expected to appear in the crystalline ground state when the Gaussian curvature is sufficient to support them. Consider therefore a putative isolated disclination of strength $q=-1$ (coordination number 7) at the waist of a catenoid. The curvature condition above requires that there exist a disk of geodesic radius $r_c$, centered on the 7-disclinaton, for which~\cite{advphysreview,pleats}  \be \int_{disk} K dA=-\frac{\pi}{3} .\ee
Clearly $r_c$ must be less than the geodesic distance $l$ from the waist to the
boundary~\cite{pleats}. For a given size catenoid $c$, we calculate $l$ and the integral of the Gaussian curvature over the geodesic disk of radius $l$. The value of $c$ for which the integrated curvature equals $-\pi/3$ is the critical value of $c$ for the appearance of  7-disclinations. We compute the integral of the Gaussian curvature numerically. We first
construct a family of geodesics radiating from the core 7-disclination (at $u=0,v=0$) by solving the geodesic equation: \be
\frac{d^2x^{\mu}}{d\lambda^2}+\Gamma^{\mu}_{\rho\delta}\frac{dx^{\rho}}{d\lambda}\frac{dx^{\delta}}{d\lambda}=0,
\ee in which $x^1=u,x^2=v$ and $\Gamma^{\mu}_{\rho\delta}$ is the
Christoffel symbol of the second kind. This second order differential equation has a unique solution given an initial position and an initial velocity. The initial conditions are
$x^1(0)=x^2(0)=0$, $\frac{dx^1}{d\lambda}\big{|_0}=\frac{1}{c}
\cos\theta$, and $\frac{dx^2}{d\lambda}\big{|_0}=\sin\theta$, where $\theta$
is the angle of the initial velocity with respect to $\vec{e}^u$. Given a geodesic radius $r$, the coordinates of the end point of the geodesic curve can be found. These end points form the
boundary of a disk in $\{ u,v\}$ coordinates (see Fig.$\ref{geodesics}$(a)). We then integrate the Gaussian curvature over the prescribed disk numerically. The critical value
of $c$ is found to be $c^*=0.85$ and the corresponding critical radius is $r_c=0.53$. Note that
integrated Gaussian curvature for this critical catenoid is quite large~\cite{pleats}: $\int KdA=-6.6$. The critical value $c^*$ can also be estimated as follows. By introducing Gaussian normal coordinates $(r,\theta)$ centered on a 7-disclination at height $z_0$ above or below the waist of the catenoid, the effective (screened) disclination charge at distance r is given by
~\cite{interactingdefectsfrozen,Frankel} \be
\rho_{eff}(r)=-\frac{\pi}{3}-\int_0^{2\pi} d\theta \int_0^{r}dr'
\sqrt{g}K(r')\\\nonumber =-\frac{\pi}{3}+\pi \frac{r^2}{c^2}
\rm{sech}^4(\frac{z_0}{c})+ {\cal O}(r^3) .\label{seff}\ee The
critical radius is reached when the effective disclination density vanishes: $\rho_{eff}(r_c)=0$. For a 7-disclnation on the waist ($z_0 = 0$) this gives $ r_c/c \equiv \theta_c =\sqrt{1/3} \approx 33^{\circ}$.
Now on the catenoid the geodesic length from the waist to the boundary is $\int^{z_m}_0 \cosh(v/c)dv = \sqrt{1 - c^2}$. The critical catenoid size $c^*$ is then given by $r_{c^*}=\sqrt{1-{c^*}^2}$.  This yields $c^*=\sqrt{3}/2\approx0.87$. This estimate for $c^*$ is very close to the numerical value 0.85. Why are these two values so close? In calculating the effective disclination charge, we use
$K(0) \pi r^2$ to approximate the integral of the Gaussian
curvature over a geodesic disk of radius $r$. The Gaussian
curvature is overestimated as its magnitude is maximum at $r=0$
(on the waist). On the other hand, since $ K(0)=lim_{r\rightarrow
0}\frac{12}{\pi r^4}(\pi r^2-A(r))<0 $, the real area $A(r)$ of
the disk with geodesic radius $r$ is bigger than $\pi r^2$, i.e.,
the disk area is underestimated in our approximation. These two
approximations tend to cancel each other out. For a typical value
of $c=1/2$, $ |K(0)| \pi r^2$ and the numerical result of the
integral of the Gaussian curvature versus $r$ is plotted in
Fig.$\ref{geodesics}$(b). As expected the flat space approximation
$K(0) \pi r^2$ is good for small $r (r<0.2)$.

\begin{figure}[htbp]
\subfigure[]{
\includegraphics[width=2.4in]{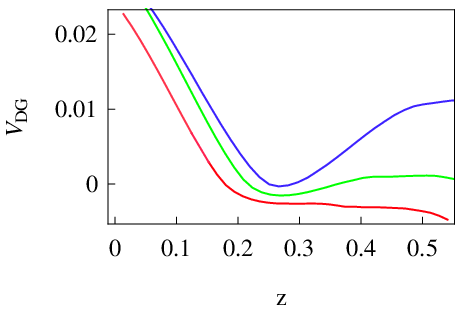}}
\subfigure[]{
\includegraphics[width=2.4in]{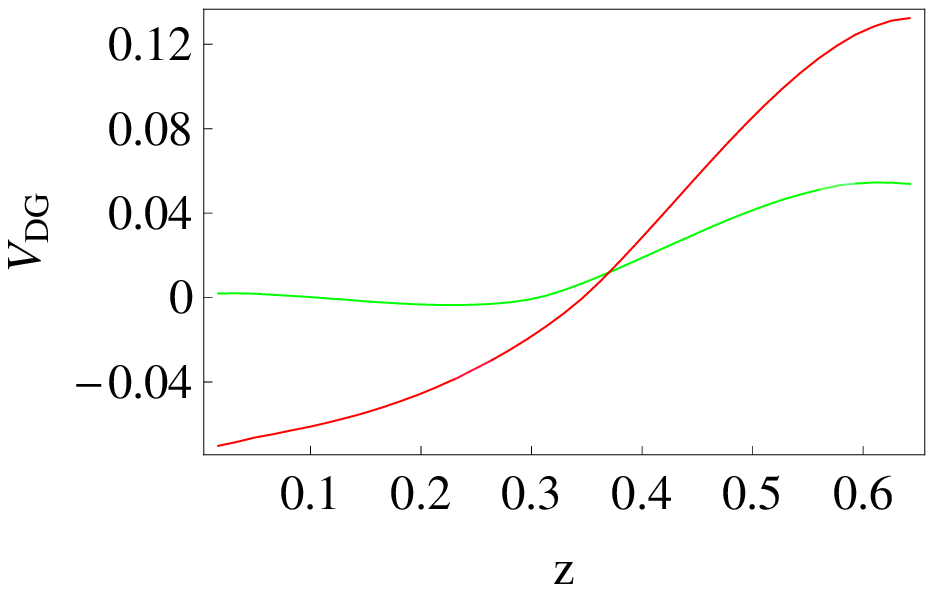}}
\caption{(a) The geometric potential of an isolated disclination for
three different values of $c$: $c = 0.8$ (red),  $c =
0.75$ (green) and  $c = 0.7$ (blue). The optimal position of an isolated
disclination moves from the boundary to the waist of the
catenoid in the rather narrow window $c$ between $0.8$ and $0.75$.
(b) The geometric potential of an isolated disclination for catenoids with $c = 0.5$ (red) and $c = 0.6$ (green).} \label{VDG}
\end{figure}

The critical waist size $c^*$ can also be estimated from energetic arguments. From the free energy of Eq.($\ref{Felchi}$) one can analyze the geometric potential describing the interaction between disclinations and the intrinsic Gaussian curvature of the surface. The result is shown in Fig.$\ref{VDG}$).
We see that the optimal position of a disclination shifts from the boundary to the waist as $c$ decreases.  The transition point for the emergence of a disclination in the interior of a catenoidal capillary bridge is
$c^*\approx 0.8$, again consistent with the value obtained above based on geometrical arguments.

\begin{figure}[htbp]
\centering
\includegraphics[width=2.3in,bb=0 0 514 319]{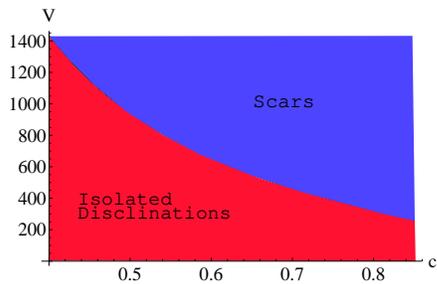}
\caption{The phase diagram in the particle number-aspect ratio plane for isolated disclinations versus scars for $c < c^*$.}
\label{phasefig}
\end{figure}

Net disclination charges may appear either in the form of
point-like isolated disclinations or extended linear grain boundary scars. Scars result from the screening of an isolated disclination by chains of dislocations and typically arise when the number of particles exceeds a threshold value beyond which the energy gained exceeds the cost of creating excess
defects~\cite{advphysreview}. Here we semi-quantitatively construct the phase diagram for isolated disclinations versus scars on a catenoidal capillary bridge characterized by the number
of particles and the aspect ratio of the catenoid $c$.

\begin{figure}[htbp]
\centering \subfigure[]{
\includegraphics[width=2.6in]{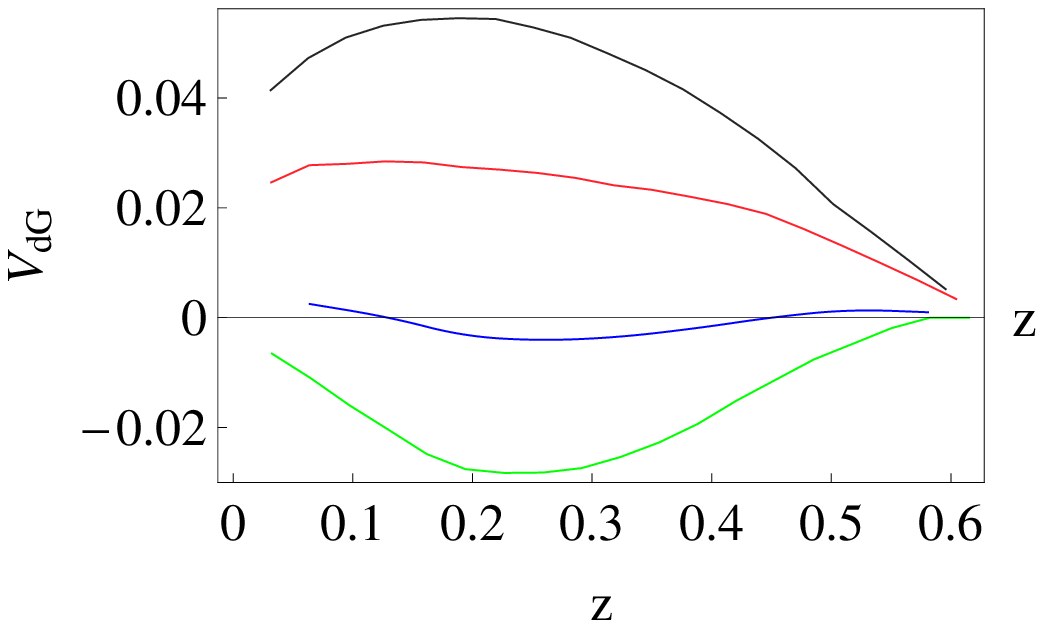}}
\subfigure[]{
\includegraphics[width=2.6in]{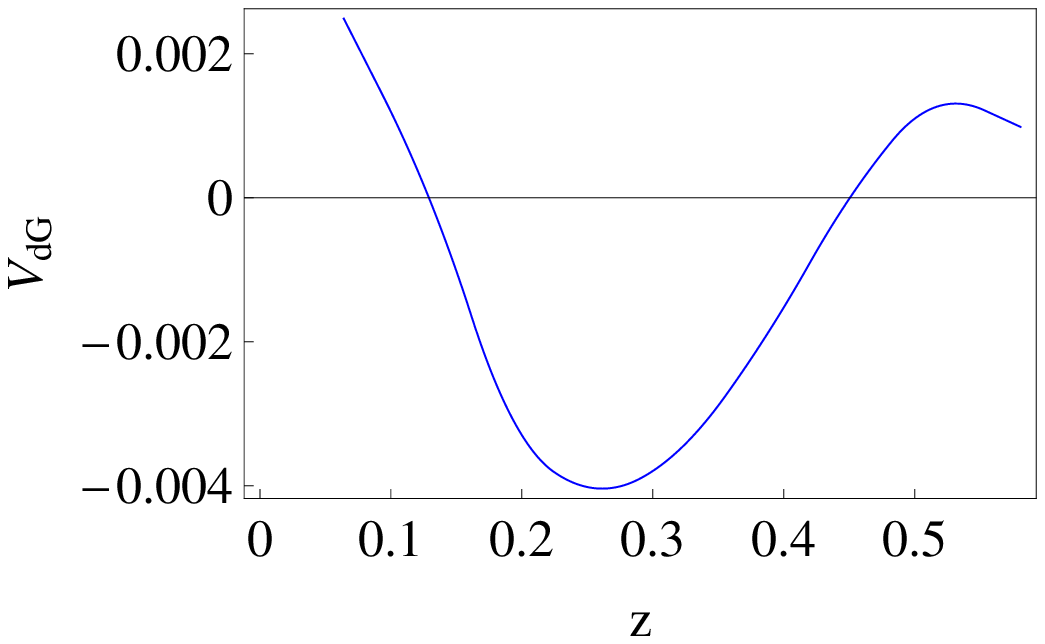}}
\caption{(a) The geometric potential of isolated dislocations as a
function of height for four different values of $c$: $c=0.7$
(black),  $c=0.68$ (red), $c=0.66$ (blue) and  $c=0.65$ (green).
The optimal position moves from near the boundary towards the
waist as $c$ decreases. (b) is an enlarged view of the blue curve
to show the transition from the red to the green curve. }
\label{dDGP}
\end{figure}

Consider a disclination on  a capillary bridge (for $c<0.85$)
radiating $m$ grain boundaries (scars).  The spacing of
neighboring dislocations is  $ l=a\
m/s_{eff}$~\cite{interactingdefectsfrozen}, where $a$ is the
lattice spacing. As $s_{eff}\rightarrow 0$, the dislocation
spacing within a scar diverges and the grain boundary terminates.
If the disclination can be completely screened by Gaussian
curvature within a circle of radius $ r \approx 3a$, then grain
boundaries will not form around the core disclination. The
condition for isolated disclinations is therefore $|K_{max}\pi
(3a)^2| \sim \pi/3$, where $|K_{max}|=1/c^2$ is the Gaussian
curvature at the waist of the bridge. On the other hand, the
number of particles $N$ is related to the surface area $A$ between
$z\in [-z_m,z_m]$ via $A(c)=\frac{\sqrt{3}}{2}a^2 N$. The curve
separating isolated disclinations from scars
 is thus given by
$ N=\frac{18\sqrt{3}\ A(c)}{c^2}$, as plotted in
Fig.$\ref{phasefig}$(a). The phase boundary reveals two basic types of transition in the topological structure of the ground state as the particle number and the geometry (aspect ratio) of the capillary bridge are varies. For a fixed catenoid aspect ratio below the critical value for the appearance of excess 7s in the interior there is a transition from isolated 7s to linear grain boundary scars with one excess 7 as the number of particles increases. For a fixed number of particles above a threshold value ($N_c \approx 300$) there is a  transition from isolated disclinations to scars as the capillary bridge gets fatter and the decreasing Gaussian curvature is insufficient to support isolated 7-disclinations.

\begin{figure}[tbp]
\centering \subfigure[]{
\includegraphics[width=2.2in]{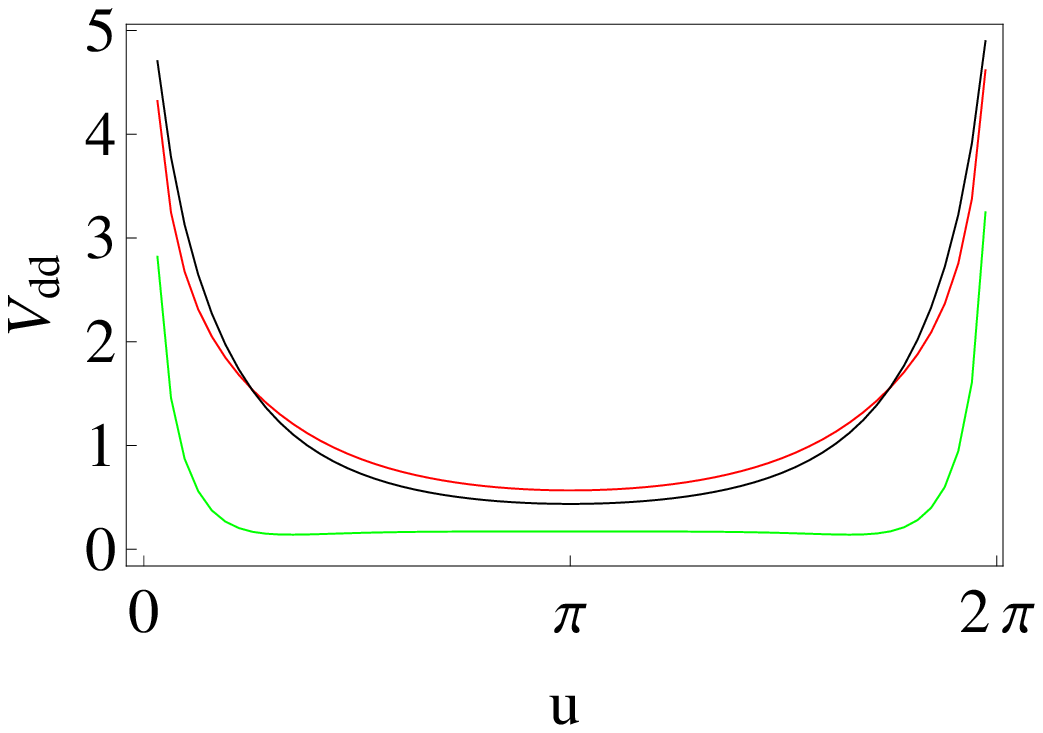}}
\subfigure[]{
\includegraphics[width=2.3in]{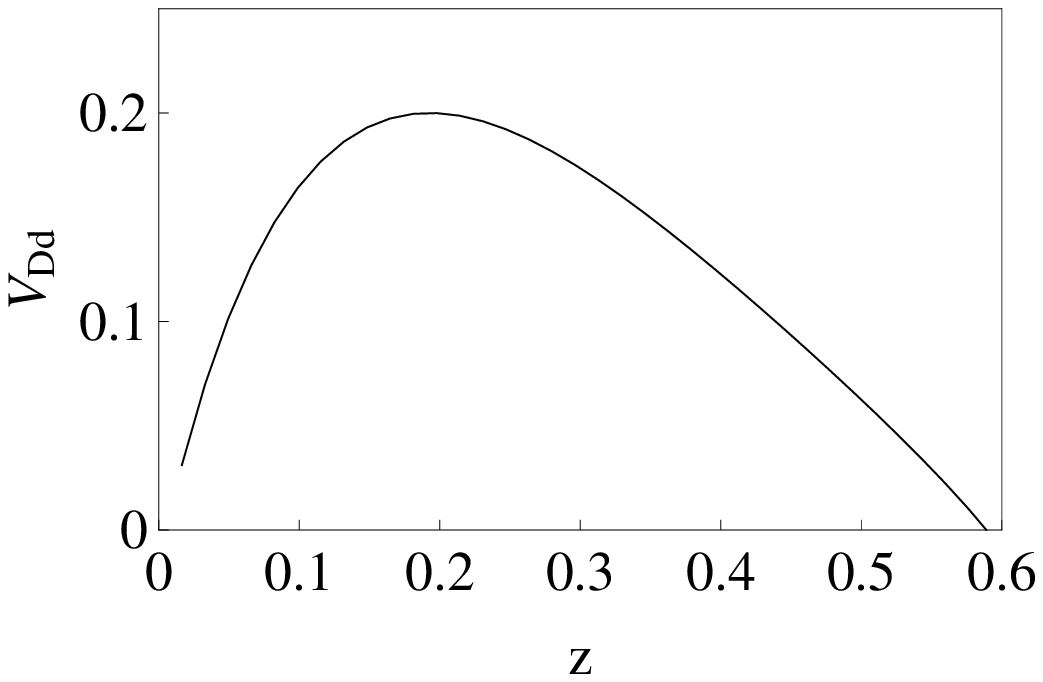}}
\caption{ The interaction of defects. (a) The
dislocation-dislocation interaction $V_{dd}$ of two dislocations
along $\vec{e}^u$ at the same height. $z=0.5\ z_m$ (black),
$z=0.3\ z_m$ (red) and   $z=0.1\ z_m$ (green). (b) The
disclination-dislocation interaction $V_{Dd}$ as a function of their
longitudinal separation. The disclination is fixed on the waist ($c=1/2$).} \label{Vint}
\end{figure}

Disclinations and anti-disclinations attract and may form dipole
bound states (7-5 pairs). Such dipole configurations are
themselves another type of point-like topological defect in
two-dimensional crystals -  dislocations. Dislocations on a
triangular lattice correspond to two semi-infinite Bragg rows
$60^{\circ}$ apart both terminating at a common point - the
location of the dislocation. Since they are tightly bound states
of disclinations the energetics of dislocations may be derived
from the governing energetics of disclinations on a curved
geometry - Eq.($\ref{Felchi}$). Dislocations, unlike
disclinations, are oriented. The Burgers vector $\vec{b}^{\alpha}$
characterizing a dislocation at position $x^{\alpha}$ is
perpendicular to the 5-7 bond. An analysis of Eq.($\ref{Felchi}$)
shows that the preferred orientation of the Burgers vector is
along $\vec{e}^u$. This is clear from the fact that the
7-disclination has minimum energy when located at the waist with
the accompanying 5-disclination in the direction of the boundary
where the negative Gaussian curvature drops most rapidly. Thus the
7-5 bond should along a meridian and the Burgers vector along a
line of latitude. From here on we restrict ourselves to this case.

The variable Gaussian curvature of a catenoidal capillary bridge
also leads to optimal positions for isolated dislocations.
Fig.\ref{dDGP} shows the geometric potential for isolated
dislocations as a function of height above the waist. As the waist
radius c decreases the optimal position of an isolated dislocation
moves from the boundary to the interior of the capillary bridge
since the increasing maximal negative Gaussian curvature
increasingly attracts 7-disclinations with their tightly bound
5-disclinations.  The boundary-to-interior transition occurs for
$c^{**} \approx 0.68$.  The corresponding integrated Gaussian
curvature  $\int KdA\approx-9.0$~\cite{pleats}. This detachment
transition is also observed in experiments with capillary
bridges~\cite{pleats} - in the experimental case the capillary
bridges are generally nodoids with non-vanishing mean curvature
and the analysis is corresponding more elaborate. The optimal
position of a single dislocation for small $c$, say $c=0.3$, is
$z(c=0.3)=0.25z_m$. Thus the optimal position of a single
dislocation is far from the waist, even for a strongly curved
catenoidal capillary bridge, in contrast to the case of
disclinations. This result can be understood in terms of the
Peach-Koehler forces acting on the individual positive and
negative disclinations that make up a
dislocation~\cite{PNAScrystallographyNelson,pleats}. While the
7-disclination prefers to be at the waist the 5 prefers to be at
the boundary - the competition results in an optimal dislocation
position somewhere in between the two extremes. Here we treat only
single dislocations but it is possible for chains of dislocations
to appear in the form of pleats, as elegantly discussed in
Ref.\cite{pleats}.

Finally we turn to the interaction between defects  themselves.
Fig.$\ref{Vint}$ (a) shows the dislocation-dislocation interaction
along $\vec{e}^u$. Two dislocations at the same height feel a
short-range repulsion. Note that near the waist shallow local
minima appear. This differs from the interaction in flat space
where parallel dislocations always repel to each other with a
logarithmic potential~\cite{ElementaryDislocationTheory}. The
attractive interaction between a disclination and a nearby
dislocation is shown in Fig.$\ref{Vint}$(b) with the disclination
fixed on the waist.

The influence of spatial curvature and topology on two-dimensional
phases of matter continues to yield surprises. The presence of
7-disclinations in negative curvature crystals may offer unique
opportunities for  functionalization of micron-scale crystallized
"superatoms" via chemistry that recognizes the unique crowded
environment of a 7-disclination~\cite{advphysreview,Stellacci}.

\section*{Acknowledgements}

We thank William Irvine, Paul Chaikin and David Nelson for discussions. This work was supported by the
National Science Foundation grant DMR-0808812 and by funds from Syracuse University.

\end{document}